\documentclass[pra,singlecolumn,superscriptaddress]{revtex4}

\usepackage{framed,multirow,threeparttable}
\usepackage{mathrsfs,amsmath,amssymb,latexsym,array,hhline}
\newcommand{\bra}[1]{\langle#1|}
\newcommand{\ket}[1]{|#1\rangle}

\usepackage{graphicx}
\usepackage{subcaption}
\newcommand{\ah}{\hat{a}}
\newcommand{\adagh}{\hat{a}^\dag}

\begin{document}
%
% Title
%

\title{The resurgence of the linear optics quantum interferometer --- recent advances \& applications}

%
% Authors
%
\author{Si-Hui Tan}\email{sihui\_tan@sutd.edu.sg}
\affiliation{Singapore University of Technology and Design, 20 Dover Drive, Singapore 138682}
\affiliation{Centre for Quantum Technologies, National University of Singapore, Block S15, 3 Science Drive 2, Singapore 117543, Singapore}

\author{Peter P.~Rohde}
\affiliation{Centre for Quantum Software \& Information (CQSI), Faculty 
of Engineering \& Information Technology, University of Technology 
Sydney, NSW 2007, Australia}

\begin{abstract}
%%%
Linear optics has seen a resurgence for applications in quantum information processing owing to its miniaturisation on-chip, and increase in production efficiency and quality of single photons. Time-bin encodings have also become feasible owing to architectural breakthroughs, and new processing capabilities. Theoretical efforts have found new ways to implement universal quantum computations with linear optics requiring less resources, and to demonstrate the capabilities of linear optics without requiring a universal optical quantum computer.%%%%
\end{abstract}

\date{\today}
%\pacs{03.67.Ac,03.67.Dd,03.67.Lx,05.40.Fb}
\maketitle

\section{Introduction}

Linear optics play a prominent role in quantum information processing \cite{bib:LovettKok}. Photons make fantastic `flying' qubits, and are readily used for quantum communication \cite{bib:Northup14} and quantum key distribution \cite{bib:Lo14}. In 2001, Knill, Laflamme and Milburn (KLM) showed that efficient quantum computing is possible using only linear optical components, that is single photons, beamsplitters, phase shifters and photon counting for spatially encoded qubits \cite{bib:KLM01}. At about the same time, Koashi, Yamamoto, and Imoto (KYI) essentially came to the same result for polarization-encoded qubits \cite{Koashi01}. Furthermore, an optical implementation of quantum protocols can potentially piggyback on existing communication infrastructure. Combined with the relative ease of transporting quantum states of light, as compared to say trapped ions or superconducting qubits, linear optics forms a powerful platform for quantum information processing.

The promises offered by linear optics have spearheaded many scientific and technical advancements. One of the major successes for linear optics is the detection of gravitational waves by the LIGO experiments based on the Michelson interferometer \cite{bib:LIGO2015}. The further application of squeezed light to the detectors promises to reduce quantum noise, which currently limits their performance \cite{bib:Barsotti18}. Another advancement is the paradigm shift from bulk optics to integrated photonics \cite{bib:Meany15}. The advantage of using such integrated photonics over bulk optics is that it is more stable against phase fluctuations, and miniaturized. This has increased the scalability of optical implementations of quantum information protocols. Furthermore, it is now possible to have single-photon sources and detectors together with linear-optical networks on a silica chip \cite{bib:Sprengers11, bib:Silverstone14}. Having all components on chip reduces coupling losses which would be crucial for fault-tolerance \cite{bib:Li15}. Experimental advancements and limitations of photonic implementations of quantum information processing are discussed at length in another review \cite{bib:Flamini2018}.

In the last decade, single photons have been produced mainly through nonlinear optical processes know as spontaneous parametric down-conversion (SPDC) and spontaneous four-wave mixing (SFWM) \cite{bib:Eisaman11,bib:Xiong16,bib:Solntsev17}. Both SPDC and SFWM produce single photons in correlated pairs that has been a staple source of entanglement. However, SPDC uses up one pump photons per pair, while SFWM requires two. Through developments in production, the number of high quality indistinguishable photons that can be simultaneously produced have increased steadily over the years. Single-photon pairs, once a novelty, are now run-of-the-mill business for quantum optical laboratories, and three or four photons have become commonplace \cite{bib:Spagnolo12, bib:Tillmann13, bib:Spring13,bib:Broome13, bib:Crespi13}. Experiments using up to ten photons have been demonstrated \cite{bib:WangChen16, bib:Chen17}. We have also seen the meteoric rise of quantum dots in the role for producing single photons \cite{bib:Ding16,bib:WangHe16, bib:213601}. They are a scalable source of single photons which are highly indistinguishable, and are produced on demand--a significant advantage over nonlinear optical processes which are inherently probabilistic.

Another approach to linear optics is to pack more features known as 
degrees of freedom onto a single photon, and to increase the degree 
of control over them. It is possible to manipulate degrees of
freedom like polarization, time-of-arrival, and orbital angular momentum \cite{bib:Tillmann2015,bib:Bozinovic2013, 
bib:Nicolas2014, bib:Humphreys2013, bib:Donohue2013}. Since KLM and KYI, new proposals 
for linear optical quantum computations (LOQC) have surfaced 
\cite{bib:Nielsen04,bib:BrowneRudolph05,bib:Gimeno-Segovia15,bib:Pant17}. 
Some of them support new functionalities like secure delegated quantum 
computations \cite{bib:Barz12,bib:Fisher14, bib:Tan16}. Others do not promise universality in their processing, but still performs tasks such as sampling \cite{bib:Aaronson11}, and encrypted quantum walks \cite{bib:Rohde_qw12} that might be hard or 
impossible with a classical system.  Such methods open up new ways to engineer 
quantum states and quantum operations, and expand the toolbox of baseline 
resources we have for implementing quantum computations, which could prove 
crucial in the race to achieve quantum supremacy.

We begin in Section \ref{sec:math} with an overview of the mathematical treatment of linear optical transformation on single photons, and in Section \ref{sec:encoding} with a description of some commonly used qubit encodings. Any unitary transformation acting on the spatial label of single photons can be efficiently decomposed into a network of beamsplitters and phase-shifters. We discuss this, and an extension to handle additional internal states, in Section \ref{sec:decomp}. For applying any linear optical circuit, one needs an accurate description of its unitary representation. A full quantum tomography is costly and time-consuming. Some practical methods have been devised to do this reconstruction without resorting to tomography, and these are described in Section \ref{sec:reconstruction} while quantum tomography is briefly discussed in Section \ref{secQPT}. In Section \ref{sec:state}, we recap the experimental breakthroughs in state preparation of single photons, and some entangled quantum states. This is followed by advances in time-bin based architecture for linear optical networks, photodectors, and optical switches in Sections \ref{sec:timebins}, \ref{sec:detector}, and \ref{sec:switch} respectively. Last, but not least, we review applications for linear optical interferometry in Section \ref{sec:applications}.

\section{Mathematical background}\label{sec:math}

A single photon in a quantum interferometer is described by its creation and annihilation operators, $\adagh_{j}$ and $\hat{a}_j$ respectively, where $j$ is the mode label of the interferometer. These operators satisfy the bosonic comutation relationship $[\ah_{j},\adagh_{k}]=\delta_{j,k}$. The action of a $2d$-port linear optical interferometer that has an equal number of input and output ports is expressed as an application of unitary operations on the creation operators,
\begin{align}
b_i^{\dag}=\sum_{j=1}^d U_{ij}a_j^\dag \ ,
\end{align}
where $a_j^\dag$ and $b_i^\dag$ are the creation operators of a single input and output photon in the $j$-th and $i$-th modes respectively, and $U\in {\rm SU}(d)$. All such transformations can be expressed as sequences of beamsplitters and phase-shifters \cite{bib:Reck1994} (see Section \ref{sec:decomp}). By convention, the interferometer is assumed to act only on the spatial mode of the input state.

Additional quantum labels are added to the creation and annihilation operators when other degrees of freedom, such as polarization, orbital angular momentum, and time-bins, are present. In this case, their commutator relation is
\begin{align}
[\ah_{j,\alpha},\adagh_{k,\beta}]=\delta_{j,k}\delta_{\alpha,\beta}\ ,
\end{align}
where $\alpha$ and $\beta$ represent these other degrees of freedom. As a consequence, quantum interference between multiple photons only occur when all quantum labels are the same. It is also possible to derive an analogous decomposition to that of Reck {\it et al.~}that realizes the unitary transformation on such photons as a sequence of beamsplitters and internal transformations \cite{bib:Dhand2015}.

Control of indistinguishability of these photons may enable future applications. Mathematical methods have been developed to deal with partial distinguishabilities among interfering photons, including those using group theory \cite{bib:Tan2013,bib:deGuise2014,bib:Shchesnovich15,bib:deGuise2015}, and quantum-to-classical transitions \cite{bib:Ra2013}. Recently, by controlling multiple degrees of freedom of single photons, genuine three-photon interference that has no independent entanglement between any two subpairs of photons was demonstrated  \cite{bib:Agne17,bib:Menssen17}.

\section{Optical encoding of quantum information on single-photons}\label{sec:encoding}

Using quantum states of light, there are a multitude of approaches to encoding quantum information. Beginning with a logical qubit,
\begin{align}
\ket\psi_L = \alpha \ket{0}_L + \beta\ket{1}_L,
\end{align}
we now discuss the most prominent such encodings, which have been widely employed. We will specifically focus on single-photon encodings, as opposed to, for example, continuous variable encodings.

These encodings are all isomorphic to one another, but nonetheless, because they are represented using entirely different physical systems, they each exhibit their own unique advantages and disadvantages, and methods by which to implement operations upon them.

%\subsection{Single-photons}

\subsection{Polarisation}

In polarisation encoding, the polarisation of a single photon in a single spatial mode encodes a logical qubit. Specifically, we represent the logical qubit as,
\begin{align}
\ket\psi_L = \alpha\ket{H} + \beta\ket{V},	
\end{align}
where $H$ ($V$) denotes a horizontally (vertically) polarised single photon.

Polarisation encoding has the elegance that the most common optical error mechanisms, such as loss or path-length mismatch, affect the two logical basis states equally. Furthermore, single-qubit operations may be directly implemented using wave-plates, which implement a rotation in polarisation space. Relevant to the preparation of large entangled states, such as cluster states, polarising beamsplitters can be employed to perform non-deterministic Bell state projections.

When physically constructing protocols based on polarisation-encoding, for obvious reasons it is extremely important that optical components be polarisation-preserving. Not doing so would obviously corrupt the logical state. Some waveguide technologies, for example, exhibit different refractive indices for the two polarisations.

\subsection{Dual-rail}

In dual-rail encoding, a single photon encodes a logical qubit as a superposition across two spatial modes,
\begin{align}\label{eq:dualrail}
\ket\psi_L = \alpha\ket{1,0} + \beta\ket{0,1},
\end{align}
where $\ket{i,j}$ is a two-mode state with $i$ ($j$) photons in the first (second) spatial mode. Using this encoding, phase-shifters and beamsplitters between the two spatial modes implement arbitrary single-qubit operations. Converting between polarisation- and dual-rail-encoding is trivial using polarising beamsplitters, which separate horizontal and vertical components into distinct spatial modes, or vice-versa.

Unfortunately, because the two basis states evolve via independent paths, our dual-rail qubits are susceptible to path-length-mismatch, a problem that does not affect polarisation encoding. Nonetheless, there are certain advantages to using two paths as opposed to one as is the case in single-rail encoding where the presence or absence of a photon denotes the logical bit. The loss of a photon in a dual-rail qubit is easily noted by its absence, whereas in a single-rail encoding, it would have been confused for one of the states. Moreover, a no-go result precludes universal quantum computation using just linear optics without the use of measurements in a single-rail encoding \cite{bib:Wu13}, thus making dual-rail encoding a more attractive alternative.

\subsection{Time-bins}

Time-bin qubits encode quantum information into the time-of-arrival of single photons, which have fixed polarisation and reside in a single spatial mode. Effectively, we discretise the direction of propagation of photons into discrete bins, which are treated as orthogonal basis states. Specifically, we are employing the encoding,
\begin{align}
\ket\psi_L = \alpha\ket{1}_{t}\ket{0}_{t+\tau} + \beta \ket{0}_{t}\ket{1}_{t+\tau},
\end{align}
where $\ket{0}_t$ ($\ket{1}_t$) denotes the vacuum (single-photon) state with arrival time $t$. Here $\tau$ is the time-bin separation, which must be sufficiently large that the temporal envelopes of neighbouring photons do not overlap, thereby ensuring orthogonality of the logical basis states.

This encoding is particularly resource-savvy, since a single spatial mode (e.g length of optical fibre) can encapsulate many time-bin qubits as a `time-bin-train'. The amount of quantum information that can be encoded into the train is limited only by its physical length.

Unlike polarisation or dual-rail encoding, time-bin encoding does not lend itself to `native' single-qubit operations. Rather, fast switching can be used to spatially separate neighbouring time-bins, implement a beamsplitter operation between them, before converting back to time-bin encoding. Typical switch times that can be achieved are in the picoseconds range \cite{bib:Dietrich16}. An experiment has built on this idea to implement a two-qubit gate by fast switching to a polarization basis \cite{bib:Humphreys2013}. Another promising advance is an ultrafast measurement technique for time bins based on converting into frequency bins \cite{bib:Donohue2013}. Scalable networks using time-bin encoding are also possible using a loop-based architecture \cite{bib:Motes14}. This is discussed in more detail in Section \ref{sec:timebins}.

\section{Efficient circuit decompositions of linear optics networks}\label{sec:decomp}

\begin{figure}[!htb]
\centering
\includegraphics[width=0.4\columnwidth]{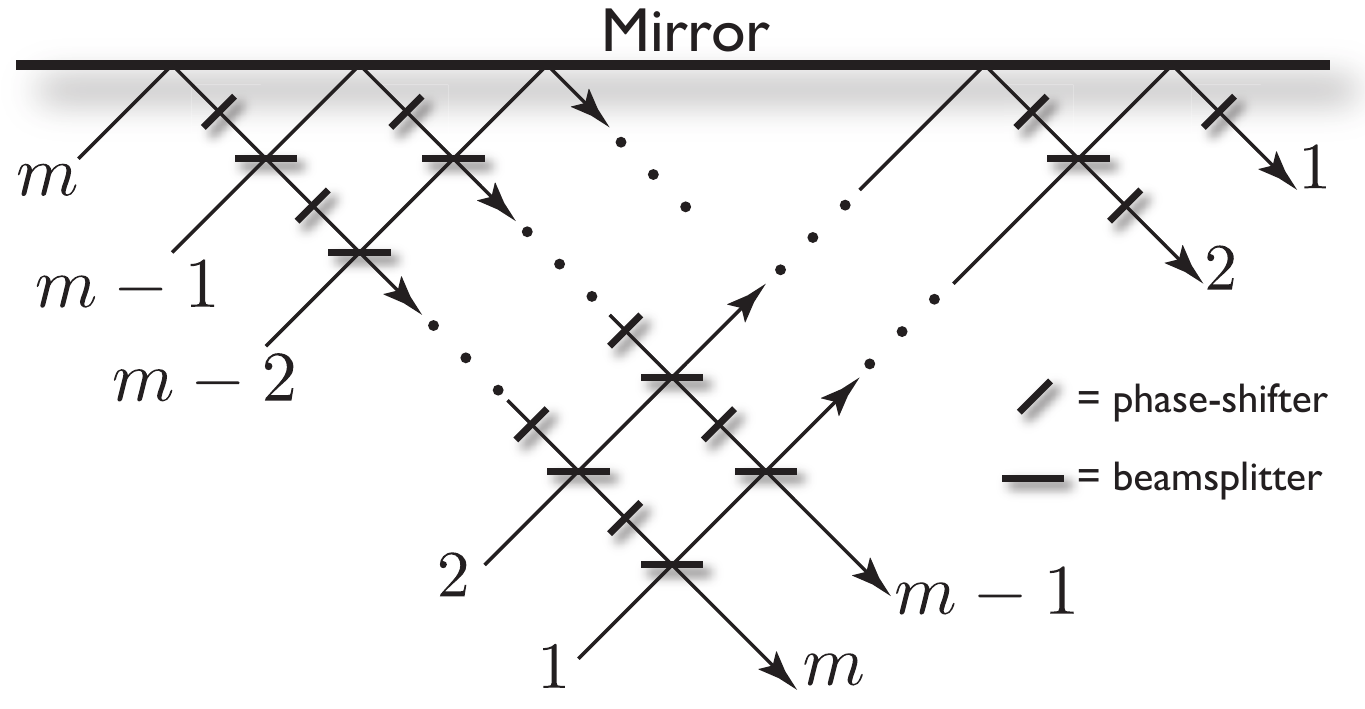}
\caption{Efficient decomposition of arbitrary linear optics 
networks into a sequence of beamsplitters and phase-shifters. This setup describes a generalized $n\times n$ multi-port Mach-Zehnder interferometer. Adapted from \cite{bib:Reck1994}.} \label{fig:Reck}	
\end{figure}

The task of implementing an arbitrary quantum computation on linear optics comes down to implementing an arbitrary $n\times n$ unitary matrix. If a non-unitary transformation is desired, it can be embedded within a unitary matrix with larger dimensions. An algorithm for expressing an arbitrary unitary matrix {\it exactly} in terms of a sequence of beamsplitters and phase-shifters was described by Reck \emph{et al.} \cite{bib:Reck1994}. This decomposition requires $\mathcal{O}(n^2)$ linear optical elements, and the algorithm for finding the decomposition has polynomial runtime. Thus, such decompositions can always be determined and implemented efficiently. The layout for the original Reck \emph{et al.} decomposition is shown in Fig.~\ref{fig:Reck}. However, since then a multitude of alternate decompositions have been found. A notable downside of the original decomposition is that different photons experience different circuit depth, i.e pass through different numbers of optical elements, resulting in asymmetry in losses and the accumulation of errors. 

\begin{figure}[!htb]
\centering
\includegraphics[width=0.8\columnwidth]{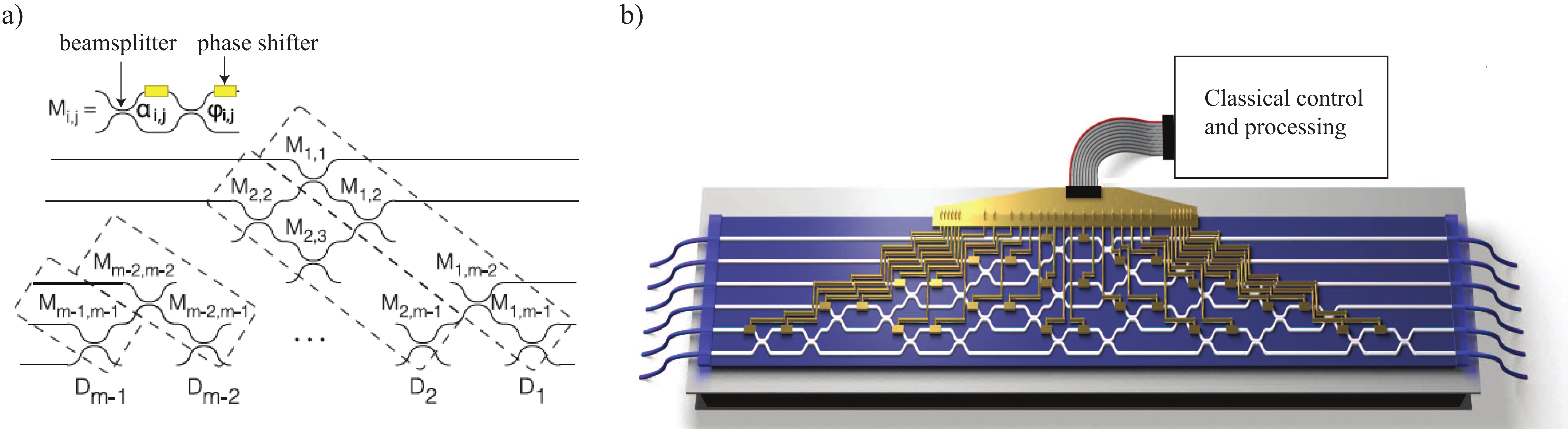}
\caption{(a) A decomposition of arbitrary linear optics networks into a sequence of subunitaries, $D_i$, made up of Mach-Zehnder interferometers, and is a universal linear-optical processing unit. These subunitaries are controlled via their photon amplitudes ($\alpha_{i,j}$) and phases ($\phi_{i,j}$). (b) A state-of-art implementation of a universal linear-optical processing unit is shown. The processor has 6 inputs entering from the left, and 6 outputs leaving to the right, and is constructed from a cascade of 15 Mach-Zehnder interferometers. The classical control and processing unit sets the parameters for the interferometer, and actively cools it. Adapted from \cite{bib:Carolan15}.} \label{fig:Carolan}	
\end{figure}

Alternatively, Mach-Zehnder interferometers can also be employed as building blocks instead of beamsplitters and phase-shifters \cite{bib:Englert2001}. Such a decomposition and its implementation on an integrated photonic chip are shown in Fig.~\ref{fig:Carolan}. Later, it was shown that any nontrivial beamsplitter, that does more than permuting modes or adding phases to them, is universal for linear optics \cite{bib:Bouland2014}. However, they do not provide an explicit construction for arbitrary unitaries.

If the linear optical transformations is to be realized on various degrees of freedom of light, then it is possible to realize a $n\times n$ arbitrary unitary transformation, where $n=n_s n_p$ for $n_s$ spatial modes, and $n_p$ internal modes, by a sequence of $\mathcal{O}(n_s^2 n_p)$ beamsplitters and $\mathcal{O}(n_s^2)$ internal transformations \cite{bib:Dhand2015}. This approach reduces the required number of beamsplitters but increases the total number of optical elements needed by a factor of 2.

\section{Reconstructing linear optics networks}\label{sec:reconstruction}

In many practical situations, the structure of a linear optical device in terms of its constituent beamsplitters and phase-shifters is known once it is built. However, owing to manufacturing imperfections, a precise characterization of these devices may still be needed post-production. One approach for achieving this is via quantum process tomography, discussed in the next section.

However, quantum process tomography is an expensive approach in terms of the number of measurements required to characterise the network, with exponential overhead, becoming impractical for large optical networks which can contain hundreds of modes using present-day technology \cite{bib:Harris16}. To mitigate this problem, alternative characterisation protocols have been developed.

Generally, the unitary matrices of $d\times d$ linear optical devices are complex $U_{ij}=r_{ij}e^{i\theta_{ij}}$, where $0 \leq r_{ij}\leq 1$, and $0\leq \theta_{ij}\leq 2\pi$. To characterise these numbers, one can do so by injecting one- and two-photon states into the network with correlated photon detection \cite{bib:Laing12}. With some mathematical simplifications, the parameters can be solved for using the one-photon transmissions, and the visibility of two-photon inputs.
%
%First, they note some equivalencies: two unitaries $U$ and $U'$ are 
%equivalent if there exist two diagonal unitary matrices $D_1^U$ and 
%$D_2^U$ such that $U'=D^U_1 U D^U_2$, because these diagonal matrices are 
%regarded as unknown and trivial phases on the input and output ports of 
%the network, to which the one- and two-photon data are insensitive to. 
%This reduces the first row and column elements to real numbers, i.e. $
%\theta_{1j}=\theta_{i1}=0$. Second, the photo-statistics remain invariant 
%under complex conjugation by $U$. Thus, the imaginary part of $M_{2,2}$ 
%must be non-negative. Then, assuming the first two rows and columns are 
%non-vanishing, and there is no total loss in the interferometer, the 
%matrix to be recovered is
%\begin{align}
%U=\left (\begin{array}{cccc}
%r_{11} & r_{12} & \ldots & r_{1m} \\
%r_{21} & r_{22}e^{i\theta_{22}} &\ldots & r_{2m}e^{i\theta_{2m}}\\
%\vdots & \vdots & \ddots & \vdots \\
%r_{m1} & r_{m2}e^{i\theta_{m2}} & \ldots & r_{mm}e^{i\theta_{mm}}
% \end{array}\right ) \ .
%\end{align}
%They showed it is possible to write the parameters  for $i,j\geq 2$  The
% remaining $2m-1$ real parameters can be found via one-photon 
%transmissions.
%
An increased accuracy in the characterization is possible by estimating 
and correcting systematic errors that arise due to mode mismatch 
\cite{Dhand16}. Others have used numerical methods to find the closest 
parameters that yield the observed visibilities 
\cite{bib:Spagnolo16,bib:Tillmann16}.

An alternative method to process tomography is feeding coherent states to probe the interferometer \cite{bib:Rahimi-Keshari13,bib:Heilmann15} instead of Fock states. Such states are produced by a standard laser source, thus reducing experimental resources. The $r_{jk}$ terms can be calculated from the ratios of output intensities at the $k$th port to the input intensity at the $j$th port. The remaining phases $\theta_{ij}$ are found by the interference pattern given by a two-mode coherent state $\ket{\alpha}\ket{\alpha e^{i\phi}}$. When the states are injected into ports 1, and $j$ respectively, the output intensity at the $k$th port is
\begin{align}
I_k=I(r_{1k}^2+r_{jk}^2+2 r_{1k}r_{kj}\cos(\phi+\theta_{jk})) \ ,
\end{align}
where $\theta_{jk}=0$ for $k=1$, and $I$ is the intensity of the input coherent states. By scanning the phase shift $\phi$ and locating the maximum value of $I_k$ for $j=2,\ldots, m$, all unknown phases can be found via $\theta_{jk}=2\pi-\phi$. An elegant modification of this scheme removes the need for precise control of the phase-shift $\phi$ by suggesting instead to plot the output intensity $I_k$ with respect to the input intensity $I$ \cite{bib:Heilmann15}. In time, the natural drift in the laser source will cause this plot to trace out an ellipse, known as a Lissajous figure, whose orientation and direction of evolution will give the phase $\theta_{jk}$ and its sign respectively.

While the input-to-output amplitudes of a linear optics network may be efficiently determined (in terms of number of measurements), it does not characterise the full quantum process it implements, which may contain errors in other degrees of freedom.

\section{Quantum state and process tomography}\label{secQPT}

When demonstrating quantum protocols experimentally it is generally absolutely critical to characterise the device prior to actual protocol execution. Tomography can be used to reconstruct the full density operator of an output state, and quantum process tomography to reconstruct the full process matrix implemented by a device. While various other measurement schemes can partially reconstruct quantum states or processes, or simply reconstruct certain parameters quantifying them, tomography is complete -- fully and uniquely reconstructing state and process matrices, from which any other measure can be determined. Importantly, state and process tomography require no interference or multi-qubit gates -- just separable single-qubit measurements in different bases. But unfortunately they require a number of measurements growing exponentially with the size of the system.

State tomography is achieved by measuring expectation values in a complete basis for the state Hilbert space. For one qubit, for example, a complete basis is the Pauli basis, which wields the completeness relatio,n
\begin{align}
\hat\rho = \text{tr}({\hat\rho}) + \text{tr}(\hat{X}\hat{\rho})\hat{X} + \text{tr}({\hat{Y}\hat\rho})\hat{Y} + \text{tr}(\hat{Z}{\hat\rho})\hat{Z}.
\end{align}
These traces correspond exactly to the expectation values in the respective bases, which are easily determined using separable measurements. Sometimes, it is possible to be able to do a tomography from an incomplete number of measurements, {\it i.e.} a number that is less than the number of expectation values, and it is possible using compressed sensing provided some sparsity conditions are met \cite{bib:GLFBE2010,bib:Titchener16,bib:Oren16}. In cases where one only wishes to know if a given state is close to a presumed state, one can perform a certification test instead of state tomography \cite{bib:Aolita15}.

Process tomography employs state tomography as a primitive, applying an equivalent completeness relation, but across the space of quantum processes, which may be decomposed into a basis input to output states. For a single qubit process this may be represented as,
\begin{align} \label{eq:process_matrix}
\mathcal{E}(\hat\rho) = \sum_{i,j=1}^4 \chi_{i,j} \hat{\sigma}_i\hat\rho\,\hat{\sigma}_j^\dag,
\end{align}
where $\hat\sigma_i$ are the Pauli operators. This generalises to an arbitrary number of qubits. Here $\chi$ is the process matrix in the Pauli basis, which fully characterises the quantum process.
%\cite{bib:Mitchell03,bib:Obrien04,bib:Lobino08,bib:Saleh11}.

%\section{Experimental implementation}\label{sec:expt}

%Significant progress has been made in the experimental realisation of 
%linear optics protocols, and have become standard and widespread. We now 
%discuss some of these advances.

\section{State preparation}\label{sec:state}

The photonic states that are most commonly employed in linear optics protocols can be divided into Fock states (e.g single-photon), and entangled states, such as EPR, GHZ and cluster states. We will consider advances in each of these. Quantum state engineering using quantum interferometers has also attracted some interest, and is discussed at the end of this section.

\subsection{Fock states}\label{sec:singlephotons}

Sources of single photons for applications in quantum information processing can be separated into two main categories: those produced by spontaneous parametric down-conversion (SPDC), and those by solid-state emitters in a cavity. To-date, SPDC has been able to produce up to ten entangled photons \cite{bib:WangChen16,bib:Chen17} that can in turn be converted into indistinguishable photons, and a quantum dot emitter in a microcavity has produced five photons \cite{bib:WangHe16}.

In SPDC, a nonlinear crystal with a large $\chi^{(2)}$ non-linearity is pumped with a laser source and with a small probability, the pump beam is absorbed by the crystal to produce two beams of lower energy known as the signal and idler. Owing to conservation of energy and momentum, the two beams have spatio-temporal correlations that can be engineered to produce twin-beam states with perfect  photon number correlation, of the form
\begin{align}
\ket\psi_\mathrm{SPDC} = \sqrt{1-\chi^2} \sum_{n=0}^\infty \chi^n \ket{n,n},	
\end{align}
where $\chi$ is the squeezing parameter. If a single photon were to be detected in one of the modes, it is certain that the other mode would similarly contain a single photon.

%Some commonly used nonlinear crystals for SPDC are beta barium borate 
%(BBO), periodically poled lithium niobate (PPLN), and periodically poled  
%potassium titanyl phosphate (PPKTP). Recently, techniques using bismuth 
%triborate (BiBO) have improved to the extent of becoming one of the record-
%holders in single-photon production \cite{bib:WangChen16}.

Solid-state single photon sources are versatile and efficient sources of single photons, however, the photons they produce have suffered from the lack of indistinguishability that is necessary for typical quantum information processing applications. Recent developments using resonant excitation of quantum dots were able to overcome these limitations. Laser pulses are used to excite the electronic resonance of the quantum dots and trigger the emission of high quality single photons \cite{bib:WangHe16}.
%\cite{bib:Ding16,bib:213601,bib:Ates09, bib:Dirk10,bib:
%1748,bib:Jayakumar13,bib:Wei14,bib:Muller14,bib:Unsleber15,
%bib:237403,bib:Sweeney14,bib:WangHe16}.

\subsection{Einstein-Podolsky-Rosen (EPR) pairs}

An EPR pair, or Bell pair, is one of the four states,
\begin{align}
\ket{0}_A\ket{0}_B \pm \ket{1}_A\ket{1}_B, \ {\rm and} \ \ket{1}_A\ket{0}_B \pm \ket{0}_A\ket{1}_B \ ,
\end{align}
which are all maximally entangled and locally equivalent to one another. These are the simplest examples of entangled states, and their preparation via SPDC has been the mainstay of entangled state preparation for quantum optical processing \cite{bib:Kim01}.
%\cite{bib:Kim01,bib:Kim03,bib:Brida07,bib:Dayan07}.
In some applications, it may be desired to have the EPR pairs conditionally prepared, {\it i.e.~}successfully prepared only under certain measurement outcomes of auxiliary modes. Because the EPR qubits are not measured directly, they can be used subsequently. Several theoretical approaches have been proposed for this purpose \cite{bib:Pittman03,bib:Sliwa03,bib:Walther07}, and demonstrated \cite{bib:Wagenknecht10,bib:Barz10}.

\subsection{NOON states}

A closely related state is the NOON state, written as
\begin{align}
\ket{\psi^{\rm (NOON)}_n}=\frac{1}{\sqrt{2}}(\ket{n,0}+\ket{0,n})\ ,
\end{align}
in the second quantization representation, and the basis is the same as that for eq.~(\ref{eq:dualrail}). This state is structurally similar to an EPR state, but its large photon-number give it great metrological power in quantum phase estimation, allowing saturation of the Heisenberg limit -- a physical bound on accuracy in phase estimation.

\subsection{Greenberger-Horne-Zeilinger (GHZ) states}

GHZ states form a class of entangled quantum states on multiple subsystems with at least three parties \cite{bib:GHZ89}. For qubit encodings with $n$ subsystems, a GHZ state is of the form
\begin{align}
\ket{\Psi_{n}^{\rm (GHZ)}}=\frac{\ket{0}^{\otimes n}+\ket{1}^{\otimes n}}{\sqrt{2}} \ .
\end{align}
These states are also non-local, and have been used extensively in experimental test for non-locality \cite{bib:JW00,bib:Zhang15}. This has been a subject matter covered in detail by another review \cite{bib:JW12}. To date, six- \cite{bib:Lu06,bib:Zhang15}, eight- \cite{bib:Huang11,bib:Yao12}, and ten-photons \cite{bib:WangChen16,bib:Chen17} GHZ states have been produced.

\subsection{Cluster states}\label{section:cluster}

\begin{figure}
\centering
\begin{subfigure}{.5\columnwidth}
  \centering
  \includegraphics[width=.5\columnwidth]{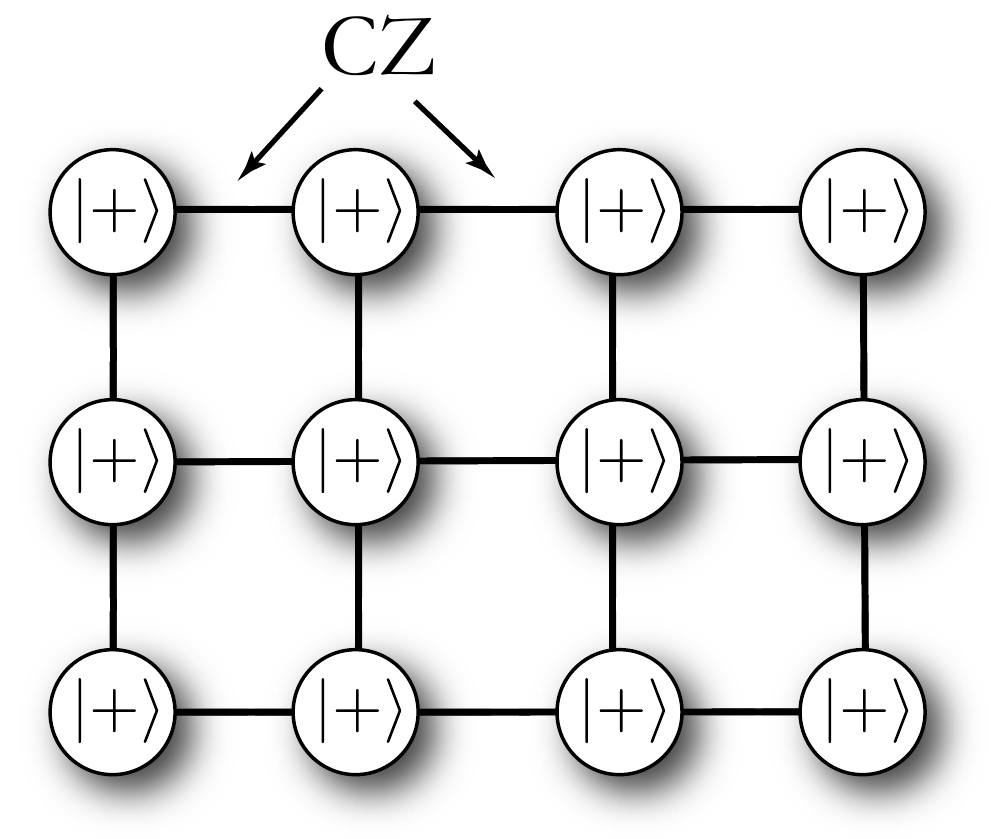}
  \caption{}
  \label{fig:cluster}
\end{subfigure}%
\begin{subfigure}{.5\columnwidth}
  \centering
  \includegraphics[width=.65\columnwidth]{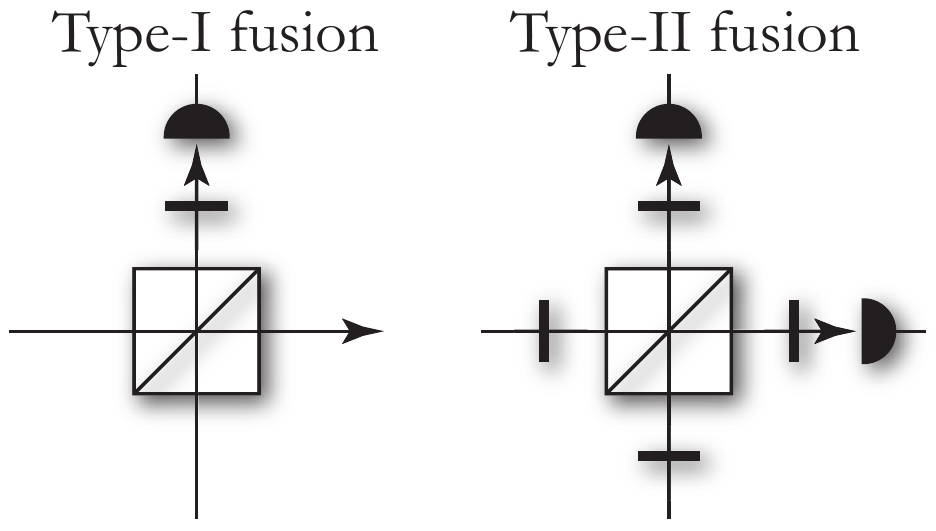}
  \caption{}
  \label{fig:fusion}
\end{subfigure}
\caption{(a) The representation of cluster states as a graph. Vertices represent qubits initialised into the $\ket{+}$ state, while edges represent the application on {\sc cphase} gates, the ordering of which is irrelevant. (b) Fusion gates for joining smaller cluster states into larger ones. Both require only a single polarising beamsplitter, and waveplates. The type-I and -II fusion gates consume 1 or 2 qubits (photons) respectively, creating an edge between the remaining graphs. The type-I gate consumes one fewer photon, but requires number-resolved detection on the detected mode. The type-II gate requires only on/off photodetection, but consumes an additional photon. Thus, the type-I gate can be employed to fuse two Bell pairs into a 3-qubit cluster state, whereas the type-II gate will only grow clusters when beginning with at least 3 qubits per cluster. }
\label{fig:test}
\end{figure}

Cluster states form another class of multiparty entangled states, and were conceived in the context of arrays of qubits with an Ising-type interaction \cite{bib:Briegel01}. These states can be represented as a graph, in which vertices are qubits initialized into the superposition state $\ket{+}=\frac{1}{\sqrt{2}}(\ket{0}+\ket{1})$, with {\sc cphase} operation applied between edges, as shown in Fig.~\ref{fig:cluster}. For this reason, such states are also referred to as graph states. The {\sc cphase} operations generate entanglement between the qubits, and because they commute, are independent of ordering.

Cluster states are a resource state for a model of quantum computation known as measurement-based quantum computation (MBQC). Here, having such a state as a resource enables universal quantum computation using only single-qubit measurements \cite{bib:Raussendorf03}. In fact, only $(X,Y)$-plane measurements are needed \cite{bib:Mantri17}. Therefore, the preparation of such states is highly valuable, generating much interest in their efficient preparation.

Unfortunately, implementing {\sc cphase} gates using linear optics is complicated and highly non-deterministic. A major improvement upon this is to use fusion gates-rotated polarising beamsplitters, which implement projections onto the Bell states. These operations fuse smaller cluster states, represented using polarisation encoding, into larger ones, consuming one (type-I fusion) or two (type-II fusion) photons in the process. These operations are shown in Fig.~\ref{fig:fusion}. Although these operations are non-deterministic with a success probability of $1/2$, they require only a single beamsplitter, already a major improvement over directly implementing {\sc cphase} gates. Furthermore, unlike {\sc cphase} gates, fusion operations require only high Hong-Ou-Mandel visibility, rather than the Mach-Zehnder stability required within existing {\sc cphase} gate implementations, a major experimental simplification.

Although these gates are non-deterministic and consume qubits, strategies have been described for efficiently preparing arbitrarily large cluster states of arbitrary topology, and using far fewer optical elements than by directly employing {\sc cphase} gates \cite{bib:Nielsen04,bib:BrowneRudolph05,bib:Gimeno-Segovia15,bib:Pant17}. Experimentally, small photonic cluster states have been demonstrated using both SPDC \cite{bib:Walther05a,bib:Lu06,bib:Prevedel07,bib:Tokunaga08} and quantum dot sources \cite{bib:Schwartz16}. To this end, continuous variable cluster states have fared better as large cluster states containing more than 10,000 entangled modes are possible  \cite{bib:Yokoyama13}, with the caveat that it is difficult to address modes individually for processing.

\subsection{Quantum state engineering based on the multi-port Mach-Zehnder interferometer}
\begin{figure}[!htb]
\centering
\includegraphics[width=0.3\columnwidth]{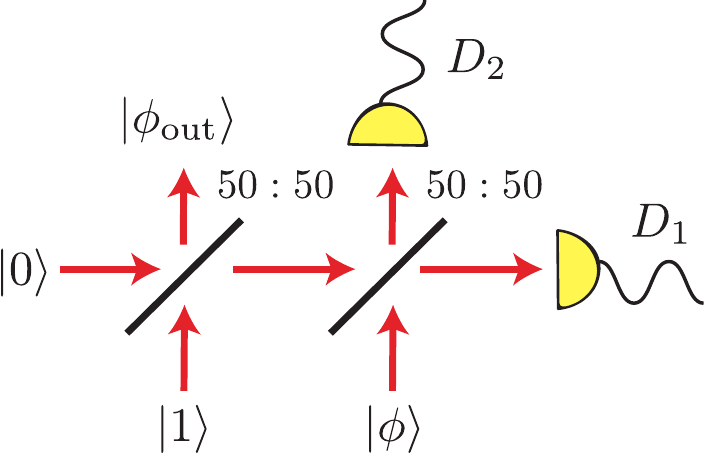}
\caption{Figure showing optical state truncation for $d=1$. The input state is $\ket{\phi}$, and when the photodetectors $D_1$ and $D_2$ register one and zero photons respectively, the output state is $\ket{\phi}_{\rm out}=c_0\ket{0}+c_1\ket{1}$ with an appropriate normalization factor. Adapted from \cite{bib:Pegg98}.} \label{fig:qscissors}	
\end{figure}

Linear optical inteferometers have been used for quantum state engineering. One well-known example is that of optical state truncation, sometimes also referred to as quantum scissors, and it happens by way of a process known as projection synthesis.  Projection synthesis works by projecting an input state onto an output state by putting the former through a multiport interferometer, and then postselecting on a certain photon number output. Optical state truncation is a use of projection synthesis to use a continuous variable state represented in the Fock-state basis, $\ket{\phi}=\sum_{n=0}^\infty c_n\ket{n}$, to produce the normalized form of $\sum_{n=0}^d c_n\ket{n}$. An example of the interferometer used for optical state truncation for the case $d=1$ is shown in Fig.~\ref{fig:qscissors}. The idea of projection synthesis was first conceived for the special case of $d=2$ \cite{bib:Pegg98, bib:Barnett99}, and later extended to $d=3$ \cite{bib:Koniorczyk2000}, and arbitrarily large $d$ \cite{bib:Miranowicz07, bib:Goyal13}. Notably, the scheme proposed in \cite{bib:Goyal13} uses multimode single photons and repeated uses of $d=2$ quantum scissors to extend the depth of truncation. 

\section{Time-bin loop-based architecture for linear optical networks}\label{sec:timebins}

\begin{figure}
\centering
\begin{subfigure}{.5\columnwidth}
  \centering
  \includegraphics[width=0.7\columnwidth]{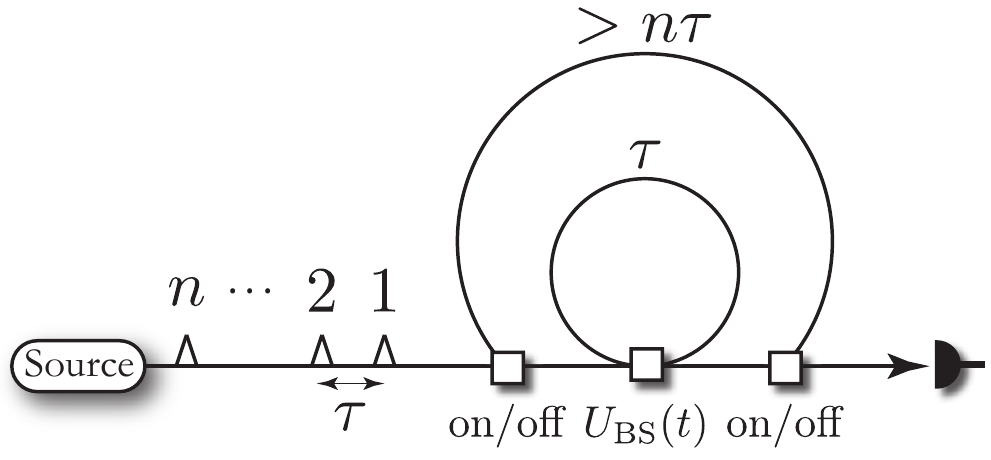}
  \caption{}
  \label{fig:FL_arch}
\end{subfigure}%
\begin{subfigure}{0.5\columnwidth}
  \centering
  \includegraphics[width=0.8\columnwidth]{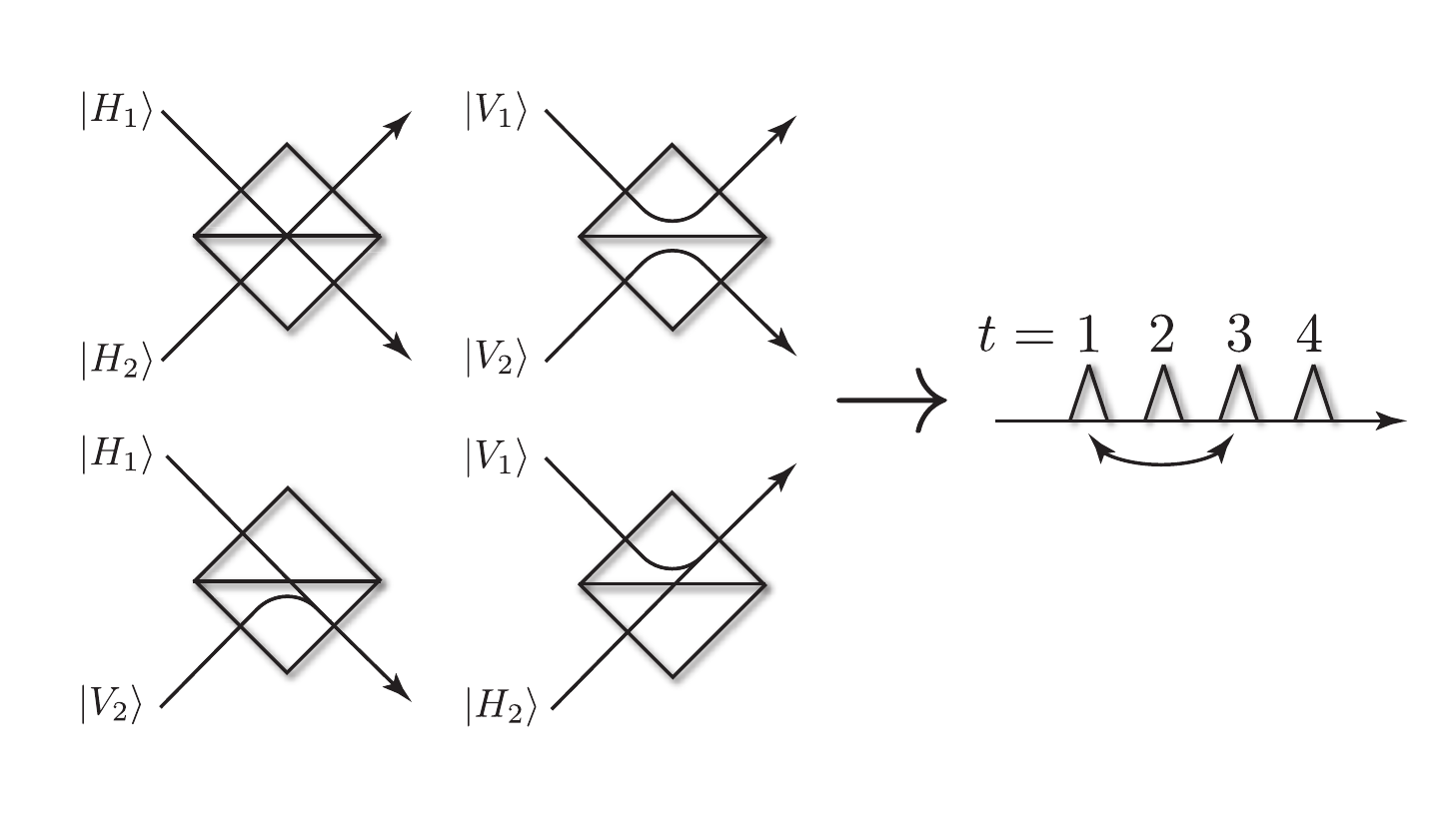}
  \caption{}
  \label{fig:PBS_TB}
\end{subfigure}
\caption{(a) A fibre-loop architecture for implementing arbitrary linear optics operations upon a time-bin-encoded pulse-train. (b) Mapping between two polarisation-encoded qubits undergoing a polarising beamsplitter (PBS) operation, and its equivalent representation using 4 time-bins in a pulse-train. The PBS completely reflects (transmits) the vertical (horizontal) polarisations. The evolution of the four logical basis states, and their respective outputs, are shown explicitly. Writing out this PBS transformation in matrix form yields a permutation. Taking this permutation and relabelling the modes, we obtain the time-bin transformation shown underneath -- a simple swap of two of the four time-bins.}
\end{figure}

When using time-bin encoding of optical qubits, the obvious question is: how do we perform single-qubit and two-qubit operations upon them. In \cite{bib:Motes14}, a dual-loop architecture was presented for implementing arbitrary passive linear-optics on photonic pulse-trains with time-bin separation $\tau$, shown in Fig.~\ref{fig:FL_arch}. The inner loop has length exactly $\tau$, while the outer one has length $>n\tau$ ($n$ is the 
number of optical modes). The architecture is controlled via three 
dynamically controlled beamsplitters. The first and last need only be on/off switches, whose sole purpose is to couple in the prepared pulse-train, 
keep it within the outer loop for the required duration, and then couple 
out of the outer loop, yielding the transformed pulse-train. The central 
beamsplitter must be able to implement arbitrary classically-controlled 
beamsplitter operations.

The architecture is frugal in its use of optical components, requiring only three dynamic beamsplitters, and several lengths of fibre. The beauty of this architecture is that the experimental requirements do not increase with the number of optical modes. The only parameter that scales with the number of optical modes is the outer loop, which must be at least long enough to house the entire time-bin-encoded pulse-train. Note, however, that the central beamsplitter must be controllable at sub-$\tau$ time-scales, so as to enable each temporal mode to be addressed individually, which is technically challenging.

The workings of the scheme can be thought of as follows: the inner loop allows arbitrary beamsplitter operations between neighbouring time-bins; the outer loop does nothing interferometric, but rather enables the pulse-train to undergo as many applications of the inner loops as necessary. It then follows that this scheme is universal for linear optics, as a sufficient number of beamsplitter operations between neighbouring modes enables universal decompositions, using for example, the Reck \emph{et al.} decomposition described in Section \ref{sec:decomp}.

As a simple example of how such time-bin encoding maps to other encodings, in Fig.~\ref{fig:PBS_TB} we show the isomorphism between the polarising beamsplitter operation and a pairwise temporal beamsplitter operation. This implies a direct mapping for implementing cluster state preparation within the time-bin scheme.

%\begin{figure}[!htb]
%\centering
%\includegraphics[width=0.6\columnwidth]{PBS_TB}
%\caption{Mapping between two polarisation-encoded qubits undergoing a 
%polarising beamsplitter (PBS) operation, and its equivalent representation 
%using 4 time-bins in a pulse-train. The PBS completely reflects (transmits) 
%the vertical (horizontal) polarisations. The evolution of the four logical 
%basis states, and their respective outputs, are shown explicitly. Writing 
%out this PBS transformation in matrix form yields a permutation. Taking 
%this 
%permutation and relabelling the modes, we obtain the time-bin 
%transformation 
%shown underneath -- a simple swap of two of the four time-bins.} 
%\label{fig:PBS_TB}	
%\end{figure}

The above description applies to passive linear optics, which is sufficient for protocols such as {\sc BosonSampling}, but insufficient for universal optical quantum computation, which requires the addition of ancillary states, and measurement with fast-feedforward. To address this, it was shown in \cite{Rohde15} that partial measurements can be implemented by coupling in and out arbitrary subsets of the optical modes. The coupling is in turn achieved via dynamically preparing ancillary pulse-trains (using the already existing source) that are classically controlled by time-resolved measurements at the output, and changing the switching sequence conditional on these measurement outcomes.

\section{Photodetection}\label{sec:detector}

\begin{table}[t]
\centering
\begin{threeparttable}
\caption{Comparison of performance between some photodetectors. In part adapted from \cite{bib:Eisaman11}.}\label{table:PNR}
\begin{tabular}{| >{\centering\arraybackslash}m{2cm} | >{\centering\arraybackslash}m{3cm} | >{\centering\arraybackslash}m{2cm} |>{\centering\arraybackslash}m{2cm} |>{\centering\arraybackslash}m{2cm} |} 
 \hline 
\vspace{0.8mm} Detector type & detector efficiency (\%) & timing jitter ($ns$) & dark counts ($1/s$) & maximum photon number \\ \hhline{|=|=|=|=|=|}
&&&& \\
APD \cite{bib:Webster2012} & 72@560$nm$ & 0.09 & 18 & 1\tnote{a} \\
&&&& \\
SNSPD \cite{bib:Takesue07} & 0.7@1550$nm$ & 0.06 & 10 & 1\tnote{a} \\
&&&&\\
  VLPC \cite{bib:Waks03}   & 85@543$nm$  & 2 & 20 000 & 5 \\
  & & & & \\
  TES \cite{bib:Lita08} & 95@1556$nm$ & 100 & - & 7\\
 TES \cite{bib:MillerNam03} & 20@1310$nm$  & 100 & $<10^{-3}$ & 15 \\
  & 20@1550$nm$ & & & \\
  TES \cite{bib:Gerrits12} & 89@1550$nm$ & - & - & 1000 \\
  &&&& \\
 Parallel SNSPD \cite{bib:Divochiy08} & 2@1300$nm$ & 0.05 & 0.15 & 4 \\
 &&&&\\
 Series SNSPD \cite{bib:Mattioli16} & 0.5@1310$nm$ & 0.116 & - & 24\\
 &&&&\\
 MPPC \cite{bib:Afek09} & 50@400$nm$ & 20 & 115 000 & 10 \\
 & 8@800$nm$&&& \\
&&&& \\
 \hline
\end{tabular}
\begin{tablenotes}
\item[a] Has no photon-number-resolving capabilities.
\end{tablenotes}
\end{threeparttable}
\end{table}

Broadly speaking, there are two main classes of photodetectors: non-photon-number-resolving (also referred to as bucket, or on/off) detectors, and photon-number-resolving (PNR) detectors. As the names suggest, non-PNR detectors only detect the presence or absence of photons in an incident beam, while PNR detectors are able to measure its number of photons. The outcomes of a PNR detector are described by the measurement projectors,
\begin{align}
\hat\Pi_n &= \ket{n}\bra{n}, \ n=0,1,\ldots \ ,
\end{align}
while those of the non-photon-number-resolving detectors are given by
\begin{align}
\hat\Pi_\text{off} = \hat\Pi_0, \ {\rm and} \ \hat\Pi_\text{on} = \hat{I} - \hat\Pi_\text{on}\ .
\end{align}
Theoretical modeling of these photodetectors can be made to include realistic effects such as detector finite efficiency, finite-number resolution, dark counts, and crosstalk \cite{bib:LeeYurtsever04, bib:Ramilli10,bib:Kalashnikov12}. These effects are typically described as projections onto mixed states \cite{bib:BARNETT1998}, and their outcomes are represented by positive-operator valued measures. Such modeling allows us to analyze these realistic effects on various quantum information protocols, see for {\it e.g.~}\cite{bib:Ozdemir2001,bib:Sperling2014, bib:Matthews16}, and update our characterization of non-classicality \cite{bib:Sperlingbinomial12,bib:Kalashnikov12b, bib:Sperling15,bib:Heilmanndc15, bib:Tan2016}.

There are many different types of photodetectors. Detailed reviews have been written about them \cite{bib:Hadfield09,bib:Eisaman11}, so we shall only outline a few important examples. A common example of a non-PNR detector is the avalanche photodiode (APD). These devices are made up of semiconductor material, usually silicon, which generate an electron that quickly generates a cascade of electrons. This avalanche of electrons generate a current that can be measured. APDs are affordable, offer low dark count rates with reasonable efficiency, and can be made to operate in a broad spectrum of wavelengths. It is the preferred detection method in quantum information protocols as they are widely available, and do not require special environments to run like some of the other detectors, as we discuss later in this section. Nonetheless, they have their limitations. A downside to using APD is their deadtime, which is a period of time following an avalanche during which the avalanche is stopped by a quenching circuit to reset the APD. No photons can be detected during the 
deadtime and contributes to photon loss in an APD. Deadtime delays for APDs range from tens of nanoseconds to $\mu$s. Superconducting nanowire single-photon detector (SNSPD) is another  type of non-PNR detector. This detector operates in its superconducting phase just below the critical density, and its phase changes back to normal when a photon is absorbed. This causes a spike in current around it, which, when detected, would indicate the absorption of the photon. This SNSPD detector tends to be faster than the APD detector with lower dark count rates. However, they require low operating temperatures of 4K maintained by special cryogenic equipment that are expensive and bulky.

PNR detectors are a wonderful addition to the linear optical toolbox. A notable example is the superconducting transition edge sensor (TES) which has an extremely high quantum efficiency. The principle it operates on is very similar to the SNSPD, but with a much higher sensitivity that allows it to detect the energy of single photons. They have low dark count rates, however, they are also quite slow, have low maximum counting rates, and, like the SNSPD detectors, are costly to operate due to their very low operating temperatures. The visible light photon counter (VLPC) is another kind of PNR detector. The way VLPC works is in principle very similar to the APD, but with an additional layer of silicon that is lightly doped with arsenic (As). Owing to the presence of the As, a single photon absorption event always create an electrical signal that is always of the same magnitude. Thus, the output electrical signal is just proportional to the number of detected photons. The VLPC can achieve a high quantum efficiency, in fact almost as high as that of the TES, and has a distinct advantage over the TES of not requiring superconducting temperatures. However, they do suffer from high dark count rate, and low speeds.
Last, PNR detectors can be created by connecting many non-PNR detectors in parallel to one another, and summing the signals at the output. For instance, like in the theoretical proposal of \cite{bib:Sperling12}. Such an approach has been successfully demonstrated with SNSPD \cite{bib:Divochiy08}, and spatially multiplexed APDs known as multi-pixel photon counters (MPPCs) \cite{bib:Afek09, bib:Chrapkiewicz14}. Characteristic experimental values for the detector efficiency, timing jitter, dark count rates, and maximum photon number of some types of photodetectors are given in Table \ref{table:PNR}.

\section{Optical switching}\label{sec:switch}

\begin{figure}[!htb]
\centering
\includegraphics[width=0.4\columnwidth]{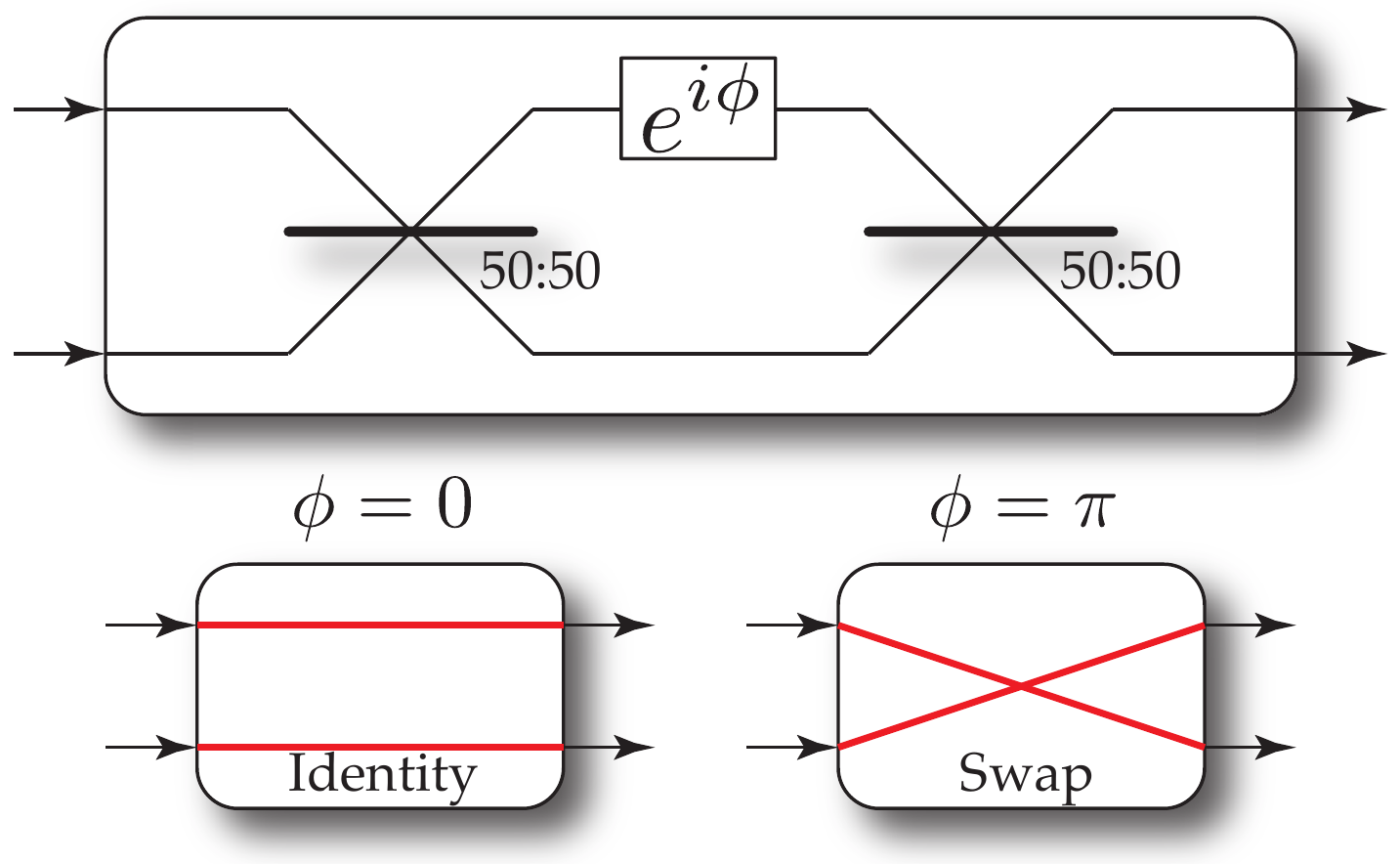}	
\caption{An optical switch for time-bin qubits that uses a Mach-Zehnder interferometer with a dynamic, classicallly-controlled phase-shift in one path.}
\label{figMZ}
\end{figure}

Some of the protocols thus far require switching between different optical modes. One way to achieve this experimentally for time-bin qubits uses a Mach-Zehnder interferometer with a dynamic, classically-controlled phase-shift in one path \cite{bib:Takesue14} (see Fig.~\ref{figMZ}). The variable phase may be implemented in a number of ways using various types of phase modulators, such as electro-optic or acousto-optic modulators. With this construction setting the phase to $\phi=0$ performs the identity operation, whereas for $\phi=\pi$ it performs a swap gate. Ref.~\cite{bib:Takesue14} reports a switching time of 100 picoseconds for this type of switch.

Another way to perform switching is to first convert temporal and spatial labels of the photon to a polarization one, and then perform a fast polarization switch before converting it back \cite{bib:HAK11,bib:Humphreys13}. Such switches are able to achieve a polarization switch within a 10 picoseconds time window. Other ways rely on conditional scattering of signal photons on ultrafast timescales, usually involving coupling between field modes and strongly correlated systems \cite{bib:Volz2012, bib:Bose2012, bib:Dietrich16}. The typical speed for this type of switching is of order of picoseconds \cite{bib:Dietrich16}.
 
This concept for switching is widely used in current laboratories, however, it is very slow using present-day technology. Although switching on its own can be completed in a timespan of the order of picoseconds, the process is slowed by the response time needed for its electronic control. Overcoming this hurdle is especially crucial in switching for fast-feedforward that is required, for example, in universal optical quantum computing. In this application, we would require total switching times, including time for control electronics to respond, on the order of nanoseconds, {\it i.e.~}faster than the time to the destination of the light it needs to switch, and this is still not possible. Presently this remains a daunting challenge, requiring much further research, with quantum memories being a candidate for loosening the switching time limitations.

\section{Applications for linear optics interferometry}\label{sec:applications}

\subsection{Linear optical quantum computation (LOQC)}\label{sec:LOQC}
The advancements in linear optics have opened up avenues for implementation. One of the key requirements is nonlinear couplings between the different optical modes. Photons do not naturally interact with one another so some careful engineering is needed to achieve this coupling. Methods used include measurements and feedforwarding \cite{bib:KLM01}, and photon-atom interactions \cite{bib:Brod16}. The former requires non-deterministic operations, which succeed only part of the time and have to be repeated. This can be challenging given the limited resources to begin with. Various linear-optical implementations of two-qubit gates important for approximately universal quantum computation are listed in Ref.~\cite{bib:Bartkowiak2010}, and comprehensive reviews have been written previously to cover this topic \cite{bib:JW12,bib:Kok05, bib:Flamini2018}, hence we will focus on recent advances.

 Integrated photonic circuits have greatly improved the feasibility of LOQC. Their small size and interferometric stability enable the requisite quantum interference to happen in certain LOQC settings. Programmable circuits that are reusable have also emerged \cite{bib:Metcalf14,bib:Carolan15}. This saves resources and lead time for producing new circuits. Combining these chips with existing higher-efficiency sources and detectors will expand their capabilities. Loss can be mitigated by moving all components, {\it i.e.~}sources and detectors, onto the same chip as the linear optical circuit \cite{bib:Sprengers11, bib:Silverstone14}.  What remains is to have fast feedforward also on chip. This might pose the greatest challenge yet for universal LOQC. The bottleneck is in the speed of modulation in the feedforward circuitry; only a handful of experiments have been able to implement adaptive feedforward for the purposes of LOQC \cite{bib:Prevedel07,bib:Xiao-Song12,bib:Mikova12,bib:Zhao14}.
  
Better single-photon sources (see Section \ref{sec:singlephotons}) are part of the solution. Improvement in control of their degrees of freedom have also led to more efficient designs of circuits \cite{bib:Zhou11,bib:Lanyon09} that enabled the optical implementation of quantum teleportation of the state of a single photon \cite{bib:Wang15}. In this way, three-qubit \cite{bib:Lanyon09,bib:Micuda13,bib:Patel16} and four-qubit \cite{bib:Starek16} gates have also been implemented using LOQC. 
 
Cross-Kerr nonlinearity is a type of nonlinearity implemented through 
 atom-photon interactions. It has been thought for some time that fundamental noise limits in atom-photon interactions will prevent any help for LOQC \cite{bib:Shapiro06,bib:Shapiro07,bib:Gea-Banacloche10}. However, this belief has been countered by the construction of a {\sc {\sc cphase}} gate using photons that undergo cross-Kerr interaction \cite{bib:Brod16}. Regardless of who is right in this debate, stronger Kerr nonlinearities have been achieved in the last 
 five years \cite{bib:Hoi13,bib:Venkataraman13,bib:Feizpour15,bib:Beck16}, 
 and this avenue remains a tantalizing possibility for deterministic 
 quantum gates.

  Although there has been tremendous progress in linear optics, it seems likely that an efficient, and fault-tolerant universal quantum computer will require a hybrid system. Linear optics is still interesting in its own right for other tasks as discussed in the remainder of this section. We also briefly mention here that single photons and waveplates are used in an already commercial application, namely, quantum key distribution (QKD). While QKD fundamentally does not require interferometry \cite{bib:BennetBrassard84,bib:Ekert91}, some QKD protocols rely on the Mach-Zehnder \cite{Bennett92, bib:Hughes2000,bib:Sasaki14,bib:Takesue15} and Franson interferometers \cite{bib:Ali-Khan07,bib:Brougham13}. Practical quantum key distribution relying on photonic technology is reviewed in \cite{bib:Scarani09,bib:Hoi-Kwong14}, and incredibly has been implemented at a distance of 1200 km by Earth-satellite transmissions \cite{bib:Liao17}. On the other hand, optimal quantum-optical cloning implementations \cite{bib:Fan2014} based on Mach-Zehnder interferometers have been applied to perform attacks on known quantum cryptographic schemes \cite{bib:Bartkiewicz2013,bib:Bartkiewicz2017}. Thus, linear optics remains fascinating platform for applications and foundation work.

\subsection{Boson-sampling}\label{sec:bosonsampling}

The difficulty in producing nonlinear operations required for universal LOQC led researchers to seek out alternative ways for demonstrating its power in the near term. {\sc Bosonsampling} is the result of such an effort. It is a restricted model of non-universal quantum computation \cite{bib:Aaronson11} that constitutes sampling the photocounts from the probability distribution of identical bosons scattered by a passive linear optical interferometer. While not universal, the {\sc bosonsampling} scheme is strongly believed to implement a classically hard task. Consider such a circuit of $m$ input (and output) modes that is injected with $n$ indistinguishable single photons such that $m\sim \mathcal{O}(n^2)$. Then, the boson-sampling task consists of generating a sample from the probability distribution of single-photon measurements at the output of the circuit. The probability of detecting $s_j$ photons at the $j$th output mode is
\begin{align}
p(s_1,\ldots,s_m)=\frac{|{\rm Per}(U_S)|^2}{s_1!\ldots s_n!} \ ,	
\end{align}
%<<<<<<< HEAD:paper.tex
where $\text{Per}()$ denotes the permanent of a matrix, $S=\{s_1,\dots,s_m\}$ denotes an output photon-number configuration, and $U_S$ is obtained from the interferometer unitary $U$ by keeping its first $n$ columns and repeating its $j$th row $s_j$ times. The appearance of the permanent, which is \#P-hard in general (a complexity class even harder than NP-hard) to compute, in this statistics contributes to the hardness of the boson-sampling problem. Note that the number of configurations $|S|=\binom{n+m-1}{n}$ in the output superposition scales exponentially with $m$. For this reason, the boson sampler does not let us \emph{calculate} matrix permanents, as this would require knowing individual amplitudes with high precision, which would require an exponential number of samples. Rather, the boson sampler samples across a distribution of permanents without actually revealing any of them to the experimenter. Nonetheless, this is a classically intractable problem. It is also known that {\sc BosonSampling} is as hard as stimulating the short time evolution of a certain Hamiltonian \cite{bib:Peropadre17}.

%>>>>>>> origin/master:(PR final edits) linear_optics_review/paper.tex

\begin{figure}[!htb]
\centering
\includegraphics[width=0.3\columnwidth]{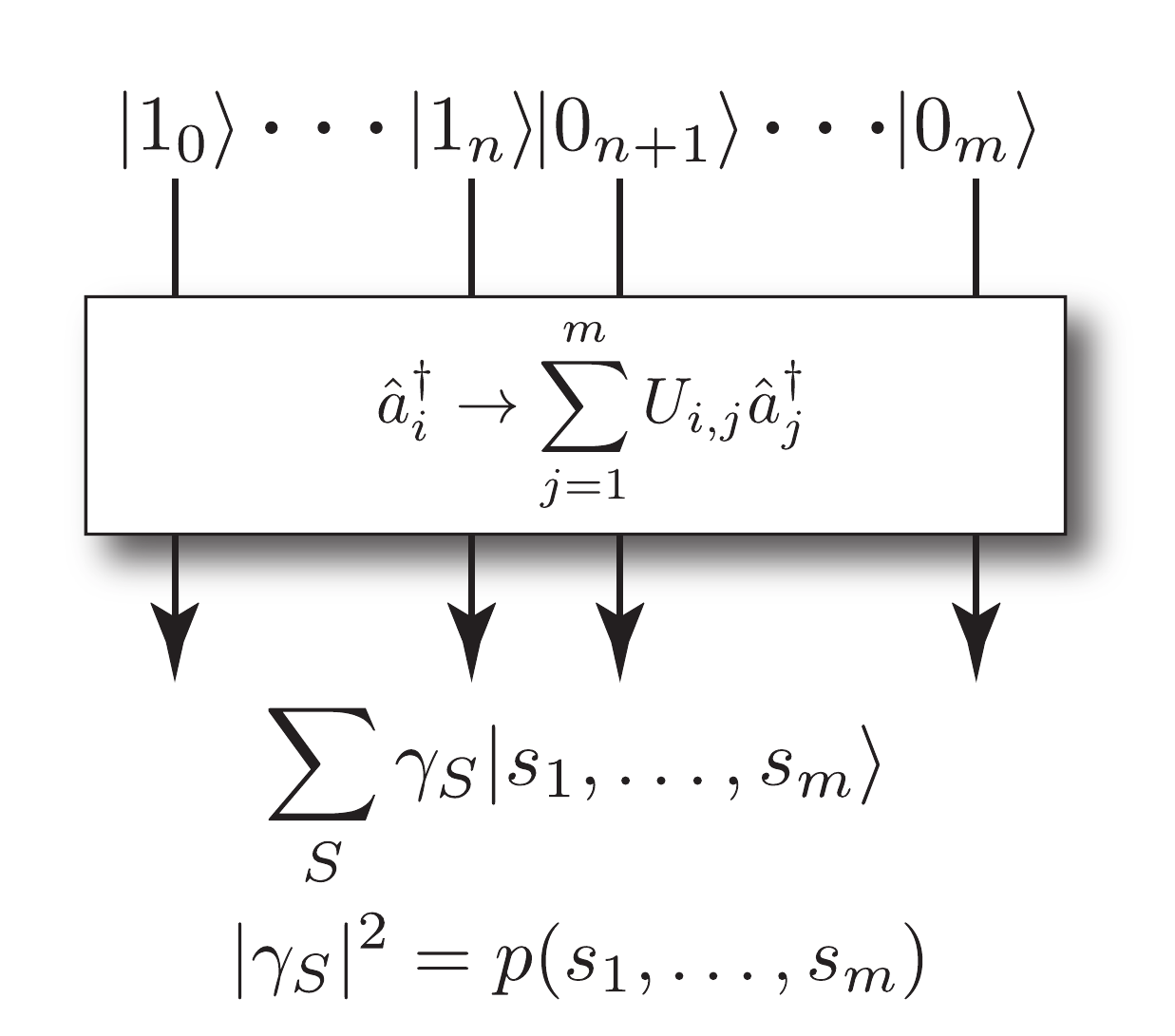}
\caption{The Boson-Sampling model. A string of $n$ single
photons is prepared in $m$ optical modes. They are evolved
via a passive interferometer $U$. Finally the photon-statistics
are sampled from the distribution $p(s_1,\ldots,s_m)$.
} \label{fig:BosonSampling}	
\end{figure} 

Owing to the simplicity of {\sc BosonSampling} and its overarching implications for computer science, multiple experiments were reported shortly after its theoretical conception \cite{bib:Tillmann13,bib:Spring13, bib:Broome13,bib:Crespi13}. A recent experiment showed a {\sc bosonsampling} machine beating an early computer \cite{bib:WangHe16}.
{\sc BosonSampling} has piqued so much interest, that a complete discussion would make up its own review, so we shall only outline some main developments here. One variant, known as scattershot {\sc bosonsampling}, proposes using pumped SPDC heralded photon sources to increase the rate of generating input photons \cite{bib:Lund14,bib:Bentivegna15}, by inputting an SPDC source into \emph{every} mode rather than just the first $n$, thereby boosting the probability of preparing the desired $n$ photons. Others varied the source input to cat-states \cite{bib:Rohde15} and thermal states \cite{bib:Tamma14_thermal} . An exciting modification allows the determination of molecular vibronic spectra \cite{bib:Huh15}, a task that goes beyond demonstrating a computationally hard task.

Certifying that a {\sc bosonsampling} task has been correctly performed is non-trivial. Some efforts look into discriminating between a valid {\sc bosonsampling} and a uniform distributions \cite{bib:Gogolin13,bib:Aaronson13}. Another shows how to discriminate data arising from either indistinguishable or distinguishable photons \cite{bib:Spagnolo14,bib:Carolan14}. More general benchmark standards have also been instituted \cite{bib:Walschaeprs16}. Another approach to verify that the interferometer is behaving as expected is to make use of suppression laws \cite{bib:Tichy14} together with fully reconfigurable optical circuits. When the circuits are tuned to implement certain unitary matrices, specific input and output combinations are suppressed \cite{bib:Crespi16}. Last but not least, a bound for the transition amplitudes for {\sc bosonsampling} indicates that an approximation algorithm for the permanent is possible \cite{bib:Yung16}. Such an algorithm can help with verifying the sampling distribution, but their proof does not construct the algorithm needed for this to happen.

Numerous other hard sampling problems have since been described, using, for example,
\begin{enumerate}
	\item Cat states (superpositions of coherent states) \cite{RohdeCat15}.
	\item Photon-added coherent states \cite{RohdeDisp15}.
	\item Photon-added or photon-subtracted squeezed vacuum states \cite{RohdePhotAdd15}.
	\item Two-mode squeezed vacuum states (SPDC states) \cite{SalehSPDC}.
\end{enumerate}

%\peter{
%Of these, 1--4 have been shown to yield hard sampling problems under photodetection, assuming certain reasonable complexity assumptions. And 5--6 have been shown to implement polynomial-time classical algorithms, under Gaussian measurements.}

As experimental imperfections threaten the scalability of {\sc bosonsampling}, there have been efforts to look into its error tolerance \cite{bib:Rohde12_tol}, fault-tolerance \cite{bib:Leverrier15}, and experimental artefacts \cite{bib:Shchesnobich14,bib:Tamma16}. Some works have resulted in interesting connections to exotic mathematical entities like immanants \cite{bib:Tan2013,bib:deGuise2014} and the multi-dimensional permanent \cite{bib:Tichy15}. A time-bin implementation of {\sc bosonsampling} \cite{bib:Motes14,bib:HeDing17} seems particularly attractive in terms of robustness \cite{bib:Motes15_timebin}. More discussions of {\sc BosonSampling} can be found in the review paper that is Ref.~\cite{bib:Lund2017}.

\subsection{Quantum simulations}
Quantum simulation involves simulating a quantum system by quantum mechanical means \cite{bib:Georgescu14}, and there are generally two approaches to achieving this. The first approach is known as digital quantum simulation and achieved by building a quantum computer from unitary gate components that are generated by a Hamiltonians of quantum processes. Progress using this approach has already been discussed in Section \ref{sec:LOQC}. The second approach is known as analog quantum simulation and involves building simpler analog devices that mimic other, perhaps less accessible, quantum systems. We focus on discussing analog quantum simulation in this section. Further discussions can be found in reviews written on quantum simulations \cite{bib:Buluta108, bib:Georgescu14}, including one specifically for optical quantum simulations \cite{bib:Aspuru-Guzik12}.

 Quantum interferometers have been proposed and in some cases implemented as quantum simulators. In quantum chemistry, the energy spectrum of the hydrogen molecule has been calculated using a quantum simulator comprising of a pair of entangled photons in a polarization-dependent interferometer \cite{bib:Lanyon10}. With two entangled photon pairs, valence-bond states have been simulated and a fiber-based tunable directional coupler gives analogue control over superpositions of different valence-bond states \cite{bib:Ma2011}. The model of {\sc Bosonsampling}, when coupled with two-level systems that are coupled to the ports of the quantum interferometer, simulates a general spin XY Hamiltonian \cite{bib:Olivares16}. Quantum interferometers can also be used to simulate quantum walks which in turn can be used to simulate topological phases \cite{bib:Kitagawa12}, transitions between fermionic and bosonic behavior \cite{bib:Schreiber11, bib:Sansoni12,bib:Matthews13,bib:Keil16}, fractional statistics of anyons \cite{bib:Han07,bib:Pachos09, bib:Chao-Yang09}, and excitation transfer in biological systems \cite{bib:Caruso11,bib:Viciani16}.
  
\subsection{Quantum metrology}\label{subsec:metrology}
\begin{figure}
\centering
\begin{subfigure}{.5\columnwidth}
  \centering
  \includegraphics[width=0.4\columnwidth]{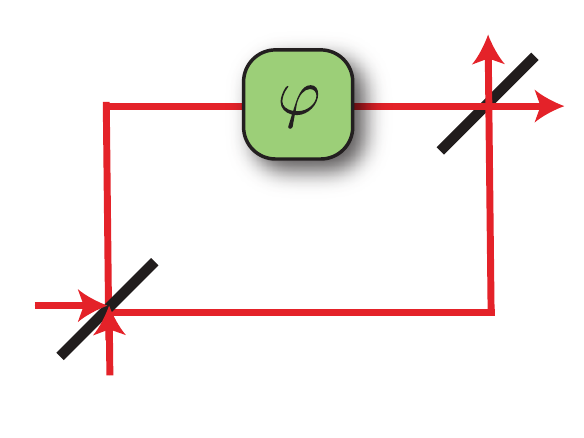}
  \caption{}
   \label{fig:MZ}
\end{subfigure}%
\begin{subfigure}{0.5\columnwidth}
  \centering
  \includegraphics[width=0.75\columnwidth]{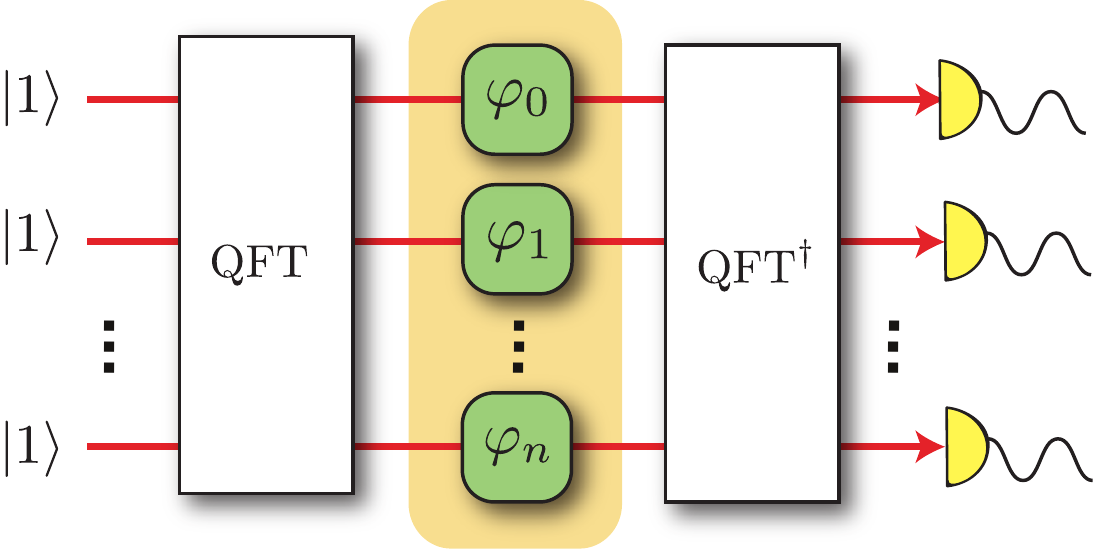}
  \caption{}
  \label{fig:MORDOR}
\end{subfigure}
\caption{(a) A schematic diagram of the Mach-Zehnder interferometer comprising of two 50/50 beamsplitters with a phase shift $\varphi$ in one of its two paths.	(b) Architecture of the quantum Fourier transform interferometer for metrology using single-photon states. QFT: Quantum Fourier transform circuit, and $\varphi_k=(k-1)\varphi$. Single photons are injected into the input modes on the left-hand-side, and a $N$-fold coincidence measurement is conducted at the output on the right-hand-side.}
\end{figure}

An archetypal task for quantum metrology is the following: Given a Mach-Zehnder interferometer, with an unknown phase shift of magnitude $\varphi$ inserted in one of the two paths (see Fig. \ref{fig:MZ}), can one deduce $\varphi$ via a judicious use of quantum states and measurements with a higher sensitivity than that of a completely classical interferometer? Indeed, one can, and the standard deviation, $\Delta\varphi$, after $N$ trials using quantum estimation is $\Delta\varphi=1/N$, which achieves a factor of $\sqrt{N}$ improvement over classical phase estimation. As this sensitivity is also a fundamental limit arising from Heisenberg's uncertainty principle, {\it i.e.} we cannot measure definitively the phase, it is known as the Heisenberg limit. Most famously, NOON states achieve this limit using $N$ photons in a single trial. Many reviews have been written for quantum metrology and Heisenberg-limited quantum phase estimation \cite{bib:Giovannetti04,bib:Dowling08,bib:Giovannetti11} so we will not dwell on the history of this application. Instead, we will discuss new paradigms for quantum metrology that have been made possible by the aforementioned advancements in linear optics.
 
% \begin{figure}[!htb]
% \centering
%\includegraphics[width=0.3\columnwidth]{phE}
%\caption{A schematic diagram of the Mach-Zehnder interferometer comprising 
%of two 50/50 beamsplitters with a phase shift $\varphi$ in one of its two 
%paths.} \label{fig:MZ}	
%\end{figure}

 As discussed in Section \ref{sec:bosonsampling}, the resources needed for Boson-sampling are readily available, but executes a task that is believed to be hard classically. Surprisingly, an entanglement between a large number of path labels is possible with single photons in a passive linear optical device. Motes {\it et al.~}harnessed this so-called number-path entanglement for quantum metrology without the use of entangled states \cite{bib:Motes15_LOQM,bib:Olsen2017}. Here, the phase shift to be estimated, $\varphi$, manifests itself as a linear phase gradient across $N$ modes (See Fig.~\ref{fig:MORDOR}). By straddling these modes among a quantum Fourier transform and its conjugate operation, it is possible to achieve $\Delta \varphi=\mathcal{O}(N^{-3/2})$ using a coincidence measurement with just a string of $N$ single photons, for up to at least $N=25$ modes. Shortly after its construction, this scheme was realized experimentally \cite{bib:Zu-En2017}. The quantum Cram\'{e}r-Rao bound of estimating $\varphi$ in such interferometers was also analytically computed \cite{bib:Chenglong2017}.
 
 Quantum multi-parameter estimation is the extension of quantum phase estimation to sense multiple phases in a multi-arm interferometer \cite{bib:Szczykulska16}. Theoretical work has shown that it is favorable to probe all phases simultaneously rather than individually \cite{bib:Baumgratz16,bib:Ciampini16}, and it remains to realize existing proposals for implementation \cite{bib:Spagnolo12}. Nonetheless, for a Heisenberg-limited sensitivity, entangled states and joint measurements across modes are likely to be required. 
 
\subsection{Repeater networks}

Quantum repeaters are devices that distribute entanglement across a network, or teleport quantum states between nodes of a network. The purpose of such a quantum network is to communicate and process quantum information over long distances. While photons serve as a useful medium for information transfer, many implementations of photonic repeater network require some kind of nonlinear interactions involving cavities for entanglement distillation \cite{Briegel98}, quantum memories \cite{Duan01}, or feedforwarding \cite{bib:Azuma15}. The seminal work of Duan et al.~ \cite{Duan01}, in particular, uses atomic ensemble as quantum memories. Single photon interference then facilitates creation of entanglement between them. However, this makes the repeater node susceptible to interferometric instability which later schemes improve upon by using two-photon interference instead \cite{Zhao07,YCZCS08,CZCS07}. Error correction codes for photon loss and dark counts \cite{bib:RHG05} have also been applied for use in repeaters \cite{bib:MHSDN10,bib:Muralidhara14,bib:Ewert16,bib:Ewert17}. Another scheme combines the use of two-photon interference and spectral multiplexing to boost high entanglement distribution rates \cite{bib:Sinclair14}.  

Without the use of repeaters, it has been shown that the rate transmission of a channel with loss without entanglement assistance has a linear fallout \cite{bib:TGW14}, and its bound is known as the Takeoka-Guha-Wilde (TGW) bound. The minimal resources required to exceed the TGW bound is a subject of active research, and so far it has been shown that the performance of the repeater schemes of Ewert-Bergmann-van Loock \cite{bib:Ewert16}, and Sinclair et al. \cite{bib:Sinclair14} are able to exceed the TGW bound. Other benchmarking tools have also been used to analyze some of these repeater schemes in detail \cite{bib:Muralidhara16}. In terms of implementation, the greatest distance that a quantum state has been teleported to-date is 1400 km \cite{bib:Ren17}, and is demonstrated between an earth-based source with a detection unit on a satellite in space. This is a culmination of many decades of research from many groups in extending this distance. If it could be made to work with other basic units of quantum repeater networks that have been implemented \cite{bib:Kimble08,bib:Sangouard11, bib:MATN15}, a large-scale, long-distance repeater network could soon be possible.

\section{Conclusion}

Linear optical interferometry has become a leading contender for the implementation of quantum information processing protocols. The preparation, manipulation and measurement of photonic quantum information is already mainstream and widely employed, with impressive experimental accuracies.

We have discussed the various ways in which quantum information can be represented photonically, how they can be manipulated and detected, and some of their leading applications.

Although there are countless physical architectures for the implementation of quantum information processing, and it is far from certain which will win `the quantum race for supremacy, optics will always find a home as the only contender for applications involving quantum communication. For this reason, the future of linear optics interferometry is a bright one, a sentiment echoed by others \cite{bib:Rudolph17}, which will find inevitable applicability in the future quantum world.

%
% Acknowledgments
%

\section*{Acknowledgments}
The authors thank A.~Laing and J.~O'Brien for their permission to use Figure 1 of \cite{bib:Carolan15}. PPR is funded by an ARC Future Fellowship (project FT160100397). This research was supported in part by the Singapore National Research Foundation under NRF Award No.~NRF-NRFF2013-01. ST acknowldeges support from the Air Force Office of Scientific Research under AOARD grant FA2386-15-1-4082.

%
% Bibliography
%

\bibliography{paper}
\bibliographystyle{apsrev4-1.bst}

%%%%%%%%%%%%%%%%%%%%%%%

\end{document}